\documentclass[sigconf,nonacm]{acmart}

\usepackage{graphicx}
\usepackage{subcaption}
\usepackage{tcolorbox}
\tcbuselibrary{skins, breakable}
\usepackage{adjustbox}

\AtBeginDocument{%
  \providecommand\BibTeX{{%
    \normalfont B\kern-0.5em{\scshape i\kern-0.25em b}\kern-0.8em\TeX}}}

\copyrightyear{2026}
\acmYear{2026}
\setcopyright{rightsretained}
\acmConference[ITiCSE 2026]{Proceedings of the 31st ACM Conference on Innovation and Technology in Computer Science Education}{July 2026}{TBD}
\acmBooktitle{Proceedings of the 31st ACM Conference on Innovation and Technology in Computer Science Education (ITiCSE 2026), July 2026, TBD}
\acmDOI{TBD}
\acmISBN{TBD}

\begin{document}

\title[Drawing Your Programs]{Drawing Your Programs: Exploring the Applications of Visual-Prompting with GenAI for Teaching and Assessment}

\begin{teaserfigure}
  \centering
  \includegraphics[
  width=\textwidth,
  clip,
  keepaspectratio,
  ]{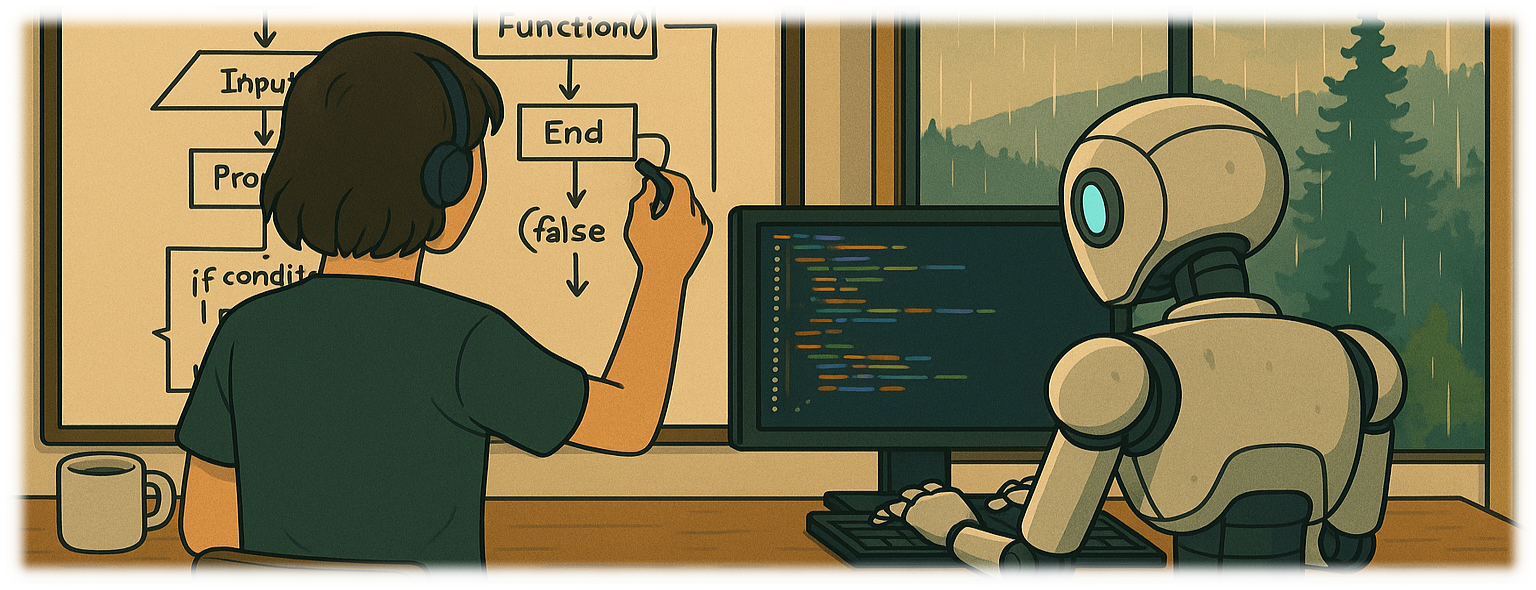}
  \vspace{-2.2em}
  \caption{Human-GenAI Collaborative Coding at a Whiteboard}
  \Description{A picture of a person sitting at a computer with a robot. The robot is coding while the person is sketching diagrams on a whiteboard. Out of the window there is a rainy PNW day and there is a cup of coffee on the desk.}
  \label{fig:teaser}
\end{teaserfigure}

\author{David H. Smith IV}
\affiliation{
  \institution{Virginia Tech}
  \city{Blacksburg, VA}
  \country{USA}
}
\email{dhsmith4@vt.edu}

\author{S. Moonwara A. Monisha}
\affiliation{
  \institution{Virginia Tech}
  \city{Blacksburg, VA}
  \country{USA}
}
\email{msheikhmoonwaraa@vt.edu}

\author{Annapurna Vadaparty}
\affiliation{
  \institution{University of California, San Diego}
  \city{La Jolla, CA}
  \country{USA}
}
\email{avadaparty@ucsd.edu}

\author{Leo Porter}
\affiliation{
  \institution{University of California, San Diego}
  \city{La Jolla, CA}
  \country{USA}
}
\email{leporter@ucsd.edu}

\author{Daniel Zingaro}
\affiliation{
  \institution{University of Toronto, Mississauga}
  \city{Toronto, ON}
  \country{Canada}
}
\email{daniel.zingaro@utoronto.ca}

\renewcommand{\shortauthors}{Smith IV et al.}

\begin{abstract}
When designing a program, both novice programmers and seasoned developers alike often sketch out—or, perhaps more famously, whiteboard—their ideas. Yet despite the introduction of natively multimodal Generative AI models, work on Human-GenAI collaborative coding has remained overwhelmingly focused on textual prompts—largely ignoring the visual and spatial representations that programmers naturally use to reason about and communicate their designs. In this proposal and position paper, we argue and provide tentative evidence that this text-centric focus overlooks other forms of prompting GenAI models, such as  problem decomposition diagrams functioning as prompts for code generation in their own right enabling new types of programming activities and assessments. To support this position, we present findings from a large introductory Python programming course, where students constructed decomposition diagrams that were used to prompt GPT-4.1 for code generation. We demonstrate that current models are very successful in their ability to generate code from student-constructed diagrams. We conclude by exploring the implications of embracing multimodal prompting for computing education, particularly in the context of assessment.
\end{abstract}

\ccsdesc[500]{Social and professional topics~Computing education}
\ccsdesc[500]{Computing methodologies~Artificial intelligence}

\keywords{Problem Decomposition, Human-AI Collaboration, GenAI}

\maketitle

\section{Introduction}
The landscape of computing is changing rapidly, with the continued rise and
evolution of generative AI (GenAI) tools and their integration into software
development practices~\cite{csimcsek2024future, vasudevan2025role,
cheng2024generative}. This shift is not only transforming how software is
developed, but also leading to reflection on the core competencies and skills across 
computing education~\cite{vadaparty2024cs1, becker2023programming,
prather2025beyond}. This, in turn, calls both for pedagogies and tools that
allow for the assessment of these skills---particularly at scale---but also
for the investigation of novel and engaging approaches to teaching them.

Chief among these skills is prompting: the ability to articulate a typically
natural language description of a task such that a GenAI model can successfully
generate code from it~\cite{denny2024prompt}. Although natural language
descriptions---whether spoken or written---are powerful tools for
communication, they are not the only ones used to represent and communicate
technical ideas. Seasoned software developers and novice programmers alike
often use diagrams~\cite{hayatpur2024taking, hungerford2004reviewing,
walny2011follow} and turn to practices such as ``whiteboard
coding''~\cite{chapin2023whiteboarding, cherubini2007let}. Many
state-of-the-art GenAI models now being natively multimodal (i.e., supporting
text, images, video, speech) opens the door to using diagrams directly as a
means of prompting. If program design often relies on diagrams, requiring that
they be translated into text for the purposes of prompting GenAI models may be
an unwelcome and unnatural redundancy---notably, departing from typical
collaborative programming practices.

In this position and proposal paper, we argue that problem decomposition
diagrams represent a promising and underexplored modality for Human-GenAI
collaboration—one that aligns with how developers naturally reason about
software design. We offer a preliminary exploration of the potential for
multimodal GenAI models---namely
GPT-4.1\footnote{https://platform.openai.com/docs/models/gpt-4.1}---to interpret
and generate code from problem decomposition diagrams provided by students. We
first articulate the affordances of visual prompting over text-based
approaches, then present a case study demonstrating that student-constructed
diagrams can effectively prompt code generation. We conclude by discussing the
pedagogical implications of this approach. In particular, if diagrams can
effectively serve as prompts for code generation, they could function as both a
visual programming interface for novices and a mechanism for providing scalable
feedback on decomposition skills.

\section{Background}
\subsection{GenAI in Computing Education}\label{sec:genai-comped}

The emergence of GenAI tools which can perform a wide range of programming
tasks, has challenged traditional expectations about what computer science
students need to understand~\cite{smith2024prompting}. As a result, there is
rising emphasis on training students to communicate intent through prompts,
critically analyze GenAI outputs, and engage in Human-AI
collaboration~\cite{lau2023ban, prather2023robots}. To understand stakeholder
perceptions, \citet{lau2023ban} interviewed CS educators from multiple
institutions and identified concerns including academic dishonesty and
degradation of fundamental programming skills from GenAI overreliance---a
sentiment echoed by students~\cite{vadaparty2024cs1}. A working group by
\citet{prather2025beyond} surveyed developers and instructors, echoing these
concerns but also recognizing opportunities to shift training toward emerging
skills like prompt engineering, testing and debugging, and Human-GenAI
collaborative problem solving. They emphasized scaffolding that enables
students to collaborate with AI tools while establishing strong computing
foundations. These findings highlight the tension between preparing students
for a Human-GenAI collaborative future and ensuring that collaboration does not
become detrimental overreliance.

\subsection{Diagramming and Sketching}\label{sec:diagramming}

Sketching and diagramming can serve as powerful tools for reasoning by, in
effect, expanding working memory by offloading its contents to a
page~\cite{kirsh2010thinking}. Similarly, they can be used as a form of
communication in collaborative design processes~\cite{tang1991findings,
goldschmidt2007see, baltes2014sketches} and for documentation of the
results~\cite{hayatpur2024taking}. In the following subsections, we offer a
brief overview of how these diagrams manifest, as well as their usage by, and
utility for, both novices and experts.

\subsubsection{Novice Diagramming and Sketching}\label{sec:novice-sketching-diagramming}

Prior work has investigated a variety of sketches used by novices
when \textit{reasoning} about programs. Sketching, particularly when tracing
through code, has been repeatedly shown to increase success in completing such
tasks~\cite{lister2004multinational, whalley2007decoding}.
\citet{cunningham2017using} performed a more recent replication of this work,
both reaffirming these findings and expanding them to understand the strategies
that were most successful for developing an understanding of the
code's behaviour. This work highlights use of such representations as a medium
for distributed cognition in the context of computing education. 

Beyond using sketches and diagrams as a means of reasoning about program
behavior, other work has investigated the use of diagramming by novices in
supporting the \textit{construction} of programs. Though this topic appears to
have received comparatively less attention, work by \citet{chapin2022effect}
suggests that the use of ``whiteboarding''---drawing out ideas in front of a
whiteboard to create or reason about programs---can play an integral role in
improving student self-efficacy, improving success at programming activities,
increasing collaboration, and facilitating an engaging learning environment.
Though this early work is promising, more work has yet to be done to better
understand and support novice programmers in both independent and collaborative
design via diagramming.

\subsubsection{Diagramming in Development Practices}\label{sec:sketching-diagramming}

Similar to novices, experts often use sketches and diagrams in a variety
of capacities. \citet{cherubini2007let} found that software developers often
employ sketches and diagrams not only to aid in their own understanding of
complex codebases but also as communicative and collaborative tools during
discussions, onboarding, and in documentation. These
diagrams, unlike those in engineering fields which make use of more formal
diagrams---are often \textit{ad hoc} due to the limited expressiveness of more
strictly formalized structures of diagramming, like UML.

\subsection{Design and Decomposition in CS Education}\label{sec:problem-decomposition}

Decomposition---though always lauded as a critical component of computational
thinking~\cite{wing2006computational}---has received increased attention as an
essential competency for students to build as they prepare to write code in the
age of GenAI-assisted programming~\cite{prather2023robots, porter2024learn,
prather2025beyond} and, as a result, one that is in need of modes by which to
assess it~\cite{srinath2025assessingproblemdecompositioncs1}. However, work
aimed at understanding the difficulties students face in designing
decompositions, as well as informing pedagogy and tools to assist in developing
this skill, appears to be sparse.

Despite this, examples of instructing students in how to engage in problem
decomposition show up most explicitly in topics related to object-oriented
programming. Seminal textbooks such as \textit{Design Patterns} instruct
students in the process of constructing programs using typical object-oriented
patterns~\cite{gamma1995elements}. In more introductory contexts,
\textit{How to Design Programs} provides an example of a text that---in addition
to teaching many of the basic constructs of programming---centers the careful
design of programs~\cite{felleisen2018design}. Perhaps the most recent example
of this includes \textit{Learn AI-assisted Python Programming: With GitHub
Copilot and ChatGPT} by \citet{porter2024learn}, which positions problem
decomposition as central to the Human-GenAI programming co-design process.

Beyond textbooks, in recent work by \citet{haldeman2025teaching}, the authors
introduce a conceptual framework for approaching program decomposition in
introductory computing courses---notably, from a perspective that heavily
leverages concepts from software engineering. They emphasize that decomposition
should be viewed from three perspectives: cohesion, reusability, and minimizing
the flow of information between components. They propose a suite of scaffolded
exercises aimed at building students' skills in adhering to these principles
when developing code.

\subsection{Prompting Activities}\label{sec:prompt-comp}

One central idea in work on CS pedagogy for the GenAI era is that we should
prepare students to \textit{communicate with GenAI models to generate code} and
\textit{evaluate the model's response to ensure its correctness}. To develop
these skills, two approaches stand out in the literature: Prompt
Problems~\cite{denny2024prompt} and Explain in Plain Language (EiPL)
questions~\cite{smith2024code, smith2024prompting}. In each, students are
provided with a representation of a computational task---a visualization for
Prompt Problems and a segment of code for EiPL---and are asked to form a
natural language description of that representation. That description is used
to generate code via a GenAI model with the description awarded marks based on
the correctness of the generated code. These approaches have shown great
promise, with students widely reporting engagement with the
activities~\cite{denny2024prompt, denny2024explaining, smith2024prompting,
prather2024interactions}, are linguistically accessible across a wide variety
of languages~\cite{prather2025breaking, li2024bridging, smith2024explain}, and
have given rise to additional autograding mechanisms~\cite{smith2025counting}
and related activities~\cite{smith2025redefining} that further bring the
approaches into alignment with their intended learning goals.

However, investigations of these approaches have been largely limited to \emph{textual}
representations of prompts. This text-centric focus is somewhat limiting from
the perspective of allowing students to engage in Human-GenAI collaboration
through all supported modalities. Our work seeks to extend prompting into the
domain of diagramming—leveraging the multimodal capabilities of modern GenAI
models to enable visual prompting for code generation. We explore this
idea---both generally as as well as how it can be situated in computing
education---via a motivating example in the following section.

\section{Motivating Examples for Visual Prompting}\label{sec:motivation}
\begin{figure*}
    \centering
    \begin{subfigure}{0.32\linewidth}
        \centering
        \includegraphics[height=150px]{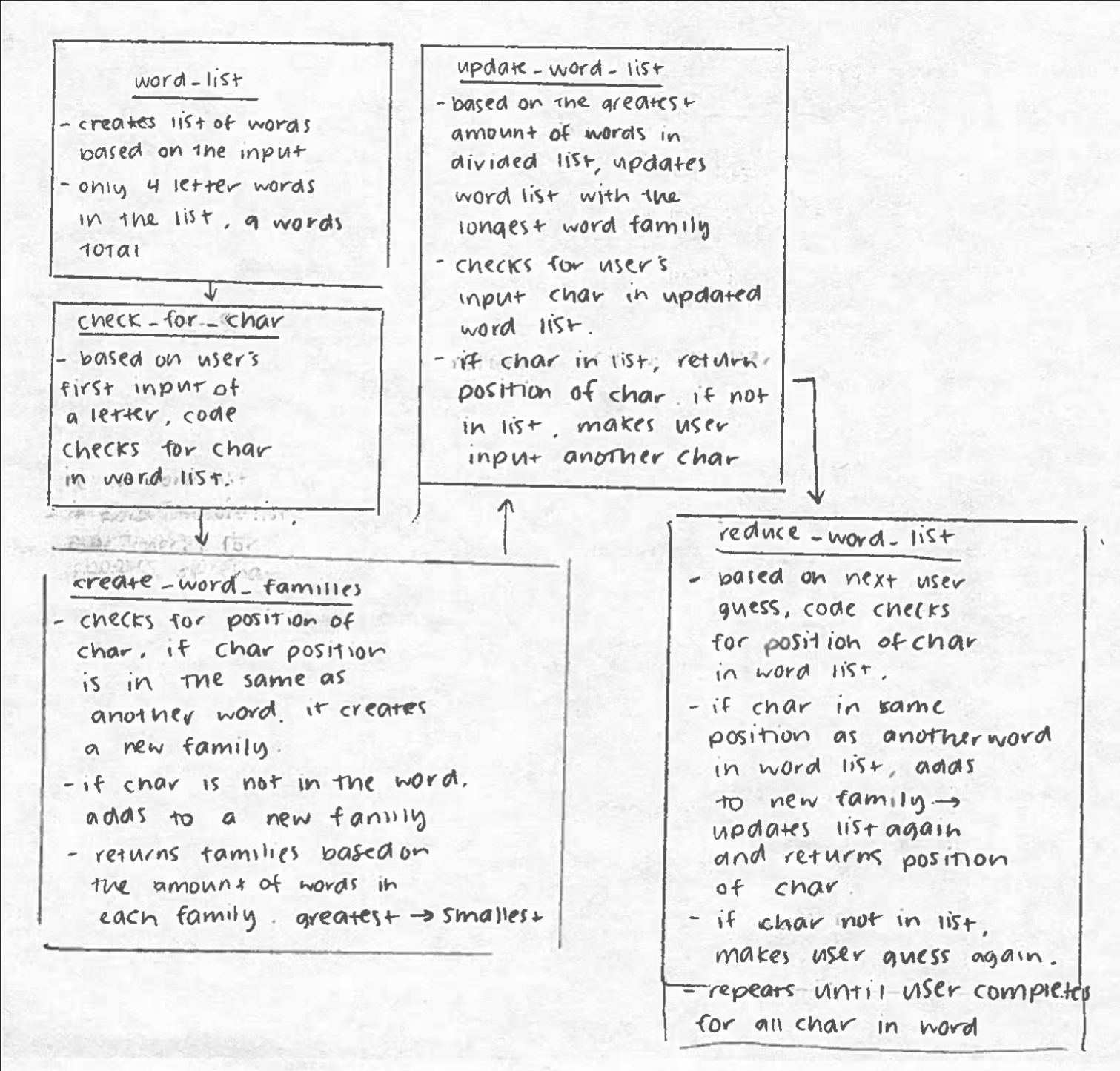}
        \caption{A simple linear decomposition diagram}
    \end{subfigure}
    \hfill
    \begin{subfigure}{0.65\linewidth}
        \centering
        \includegraphics[height=150px]{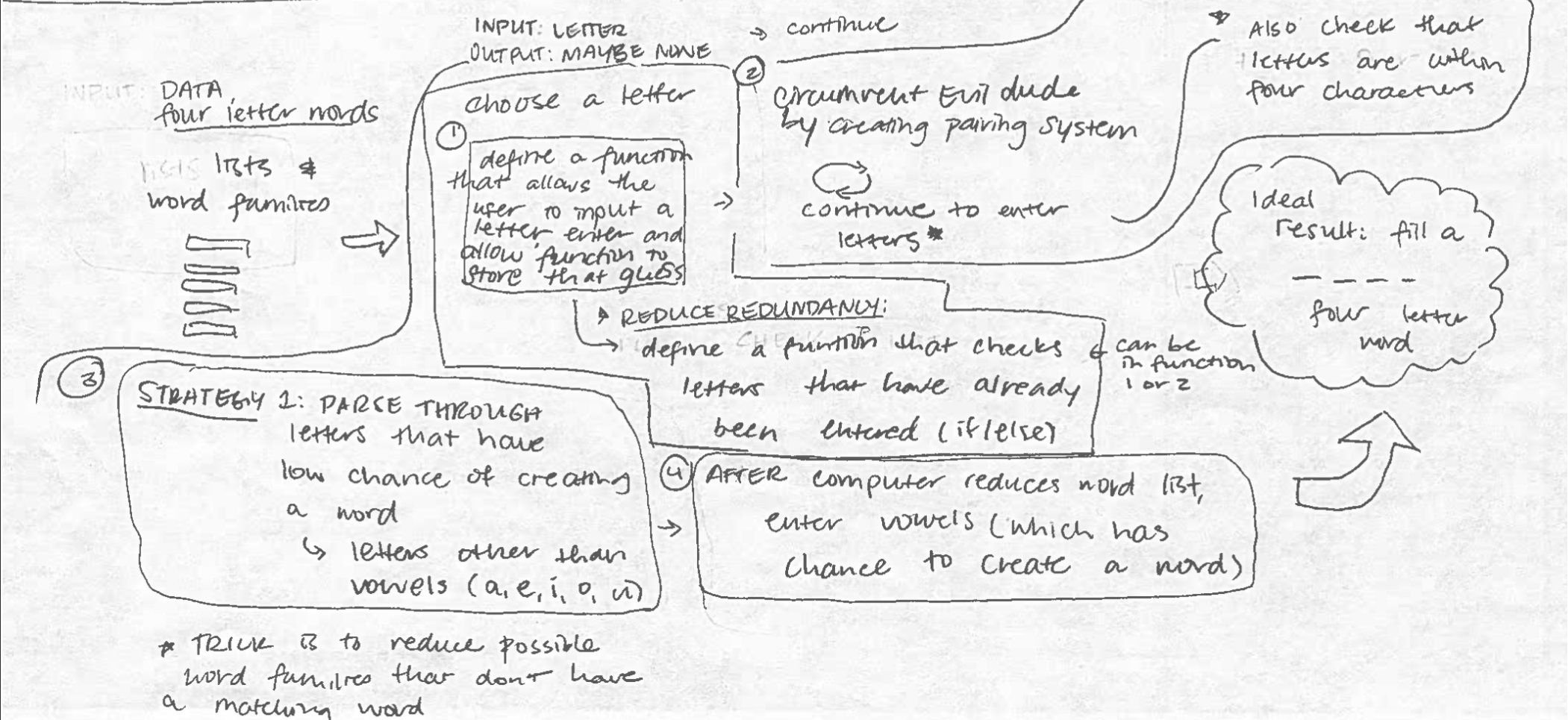}
        \caption{A more complex decomposition diagram with annotations and additional symbols}
    \end{subfigure}
    \vspace{-0.25cm}
    \caption{Two examples of program decomposition diagrams collected from the dataset.}\label{fig:diagrams}
    \vspace{-0.25cm}
\end{figure*}

While textual prompting has dominated research on Human-GenAI collaboration for
code generation, particularly in the context of computing education, we argue
that exploring the space of \textit{visual prompting} represents a promising and
underexplored avenue with particularly exciting affordances and applications in
the context of computing education.

\subsection{Authenticity of Visual Representations}\label{sec:authenticity}

As discussed in the background, perhaps the best argument for the use of 
visual prompting via diagramming lies in its authenticity. Both novice and
experts alike often use sketches and diagrams as part of their natural workflows
to explore and communicate program design~\cite{cherubini2007let, cunningham2017using}. This is, perhaps, unsurprising
as diagramming and sketching can serve both as powerful tools for reasoning
and are a natural form by which to communicate relationships and structures among
components. 

Certain aspects of program design are inherently spatial or relational and may
be more naturally expressed through visual rather than textual means. 
\begin{itemize}
    \item \textbf{Dependencies:} Which functions call which other functions can be shown through arrows or connecting lines
    \item \textbf{Hierarchies:} Parent-child relationships between components can be represented through nesting or vertical arrangement
    \item \textbf{Data flow:} How data moves through a program can be traced visually
    \item \textbf{Parallelism:} Concurrent or independent operations can be shown through horizontal arrangement
\end{itemize}
While these relationships can be described in text, doing so often requires
verbose explanations that may be less immediately clear than their visual
counterparts.

\subsection{Reduced Extraneous Cognitive Load}\label{sec:cognitive-load}

In the context of computing education, the issue of cognitive load is of
particular importance as, when students are in the process of learning a new
skill, the management of extraneous cognitive load can be critical to ensuring
students are not overloaded when engaging with the
activity~\cite{garner2002reducing,duranCLT2022}. Related to the section above,
when students are constructing a program consisting of multiple functions or
perhaps even multiple files, it is often most authentic for them to do so using
diagrams which also act as a form of distributed
cognition~\cite{cunningham2017using}.

When moving from the diagram to the process of constructing the program via
prompting, the translation from visual representation to textual description
is both inefficient and may undermine the affordances of diagramming in the
first place. Consider the workflow comparison:
\begin{itemize}
    \item \textbf{Text-based prompting:} Mental model $\rightarrow$ Diagram (optional) $\rightarrow$ Text description $\rightarrow$ Code
    \item \textbf{Visual prompting:} Mental model $\rightarrow$ Diagram $\rightarrow$ Code
\end{itemize}
By removing the intermediate translation to text, visual prompting reduces the
number of representational transformations required. This may be particularly
beneficial for novice programmers, who already face substantial cognitive
demands when learning to program and may struggle to articulate their
designs---which will likely rely on visual cues such as boxes, arrows, nesting,
etc---in precise natural language.

\subsection{New Forms of Assessment and Activities}\label{sec:new-assessment}

Finally, and perhaps most excitingly, visual prompting opens the door to new
forms of assessment and learning activities. If students can construct
diagrams, and those diagrams can be evaluated with or used to generate code via
multimodal GenAI models, this gives way to new forms of assessment that may
rely heavily on diagrams and new forms of activities that can engage students
in novel ways by allowing them to effectively ``draw'' programs. 

\subsubsection{Diagramming Assessment}

In summative contexts, similar to the autograding procedure used for both Prompt problems~\cite{denny2024prompt} and
Explain in Plain Language questions~\cite{smith2024prompting}, students could provide their diagram representing a proposed solution to a given problem. That diagram could then be used to generate code for that solution and that code could be tested to ensure its functionality meets the question author's expectations. Importantly, because students define their own functions and program structure, such assessments would have to rely on end-to-end behavioral testing rather than a simple suite of unit tests targeting specific functions. Such an approach would allow for the at-scale automated assessment of a wide variety of diagram types.

\begin{enumerate}
    \item \textbf{Entity-Relationship Diagrams (ERDs):} Students construct ERDs that are used to generate database schemas. The resulting schema can be tested against queries that verify it correctly represents the specified domain.
    \item \textbf{Program Decomposition Diagrams:} Students construct diagrams depicting function structure and relationships. The generated code can be evaluated against behavioral test cases that verify the program meets its specification (e.g., verifying the win condition of a game).
    \item \textbf{UML Class Diagrams:} Students design class structures that are used to generate object-oriented code. Test harnesses can then verify that instantiated objects exhibit expected behaviors.
\end{enumerate}

In each case, the diagram functions as a specification---a visual prompt---and
the quality of that specification is assessed through the behavioral correctness
of the code it produces. This shifts the assessment focus from the manual
production of code to problem specification, aligning with emerging perspectives
on essential skills for Human-GenAI collaborative
programming~\cite{porter2024learn}. 

\subsubsection{Program Sketching Activities}

Beyond pure assessment, code generation from diagrams enables a new class of
formative learning activities centered on iterative design. Students can sketch
a decomposition for a given task, observe the code it generates, execute that
code against provided test cases, and refine their diagram until the generated
program meets the behavioral specification. This mirrors the iterative
refinement process observed in Prompt Problems~\cite{denny2024prompt} and
Explain in Plain Language questions~\cite{smith2024prompting} but operates in a
visual modality that may be more suitable for articulating program structure.
The feedback loop of ``sketch, generate, test, revise'' could aid students in
developing an intuition for how representations of design decisions ultimately
manifest as code and behaviors, thus acting as a bridge for understanding the
connection between abstract representation and concrete implementation.

\section{Case Study}

To explore the viability of diagrams as a form of visual prompting, we conducted
a study examining student program decomposition diagrams and their effectiveness
in generating functional programs. This case study provides preliminary
empirical support for the viability of this approach and supports the notion
that further work should be done to explore the design of pedagogical tools and
practices that leverage visual prompting via diagrams. 

\subsection{Data Collection}

We conducted a study in a large introductory computer science course at a large
research university in the United States. In total, 133 students participated in
this study. Collection and analysis of this data was approved by the
university's Institutional Review Board (IRB).

As an in-person course assignment, students were given a detailed description
and worked example of the game ``Evil Hangman''---a variant of the game hangman
where the system, rather than picking a single word and having the user guess it
one letter at a time, iteratively refines a set of possible words. This set of
words is always consistent with the user's guesses, and is kept as large as
possible after each guess to make it difficult for the user to reduce this set
to a single word.\footnote{http://nifty.stanford.edu/2011/schwarz-evil-hangman/}
Students were then given fifty minutes to construct a program
decomposition diagram
for the implementation of this game on a single letter sized
(8.5x11in/21.5cmx28cm) piece of paper. Two examples of decomposition diagrams 
given by students are shown in Figure~\ref{fig:diagrams}.

We then used GPT-4.1 to generate code from these diagrams, prompting the
model with only the diagram image and minimal instructions to implement ``a
program'' based on the depicted decomposition—providing no external context
about the Evil Hangman task itself or even that it was a word guessing game.

\subsection{Analysis}

To evaluate function extraction quality, two researchers independently coded
each diagram for the number of functions present (inter-rater reliability:
$r=0.97$), then reconciled disagreements to reach consensus. We compared both
the quantity and content of functions extracted by each prompt type against
this ground truth, categorizing outputs as \textit{Fully}, \textit{Partially},
or \textit{Failed} based on whether the generated function names and docstrings
were in alignment with those defined in the diagrams.

\subsection{Results}

\begin{figure}
  \centering
  \includegraphics[width=\columnwidth]{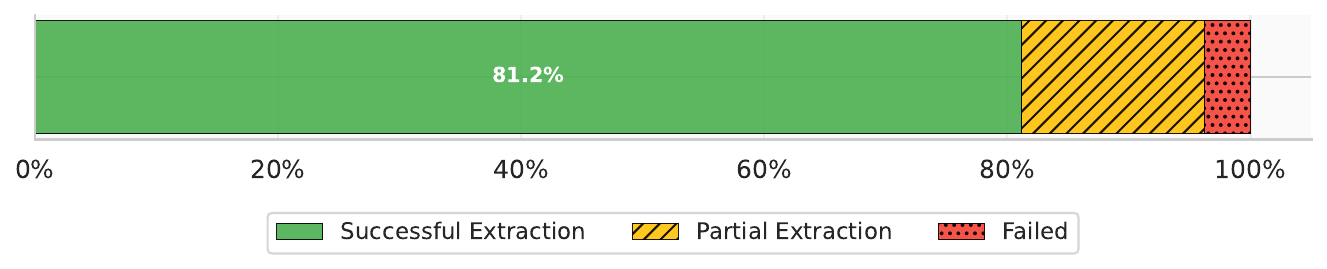}
  \vspace{-0.5cm}
  \caption{Proportion of successful versus only partially successful function extractions.}\label{fig:extraction}
  \vspace{-0.3cm}
\end{figure}

The GenAI model showed strong ability to extract functions from
student-generated diagrams, achieving a correlation of $r = 0.776$ ($p < 0.001$)
between the number of human-identified functions and the number of functions
generated. The model achieved an 81.2\% success
rate in fully extracting the function structure defined in student diagrams
(Figure~\ref{fig:extraction}).

\begin{figure}
  \centering
  \includegraphics[width=\columnwidth]{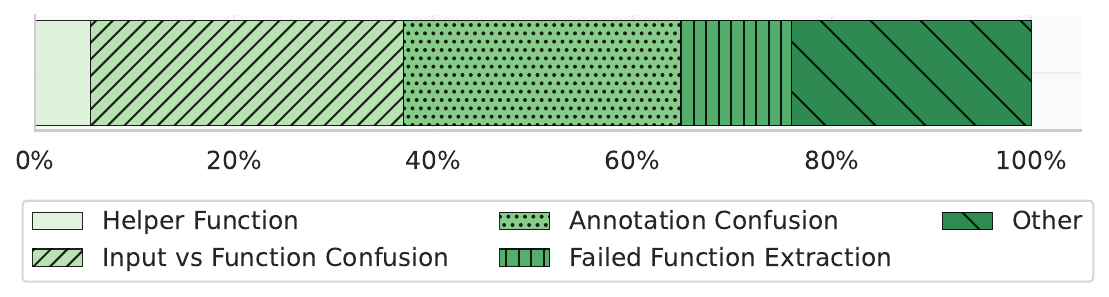}
  \vspace{-0.5cm}
  \caption{Sources of code deviations and errors linked to diagram interpretation.}\label{fig:errors}
  \vspace{-0.3cm}
\end{figure}

When deviations occurred, models consistently generated more functions than
human raters identified (Figure~\ref{fig:errors}). To understand these
deviations, we coded functions by source of deviation and explore the most
common reasons below.\\

\begin{figure}[t]
  \centering
  \begin{subfigure}[b]{0.39\columnwidth}
    \centering
    \adjustbox{valign=l}{\hspace{-1em}\includegraphics[width=1.2\textwidth]{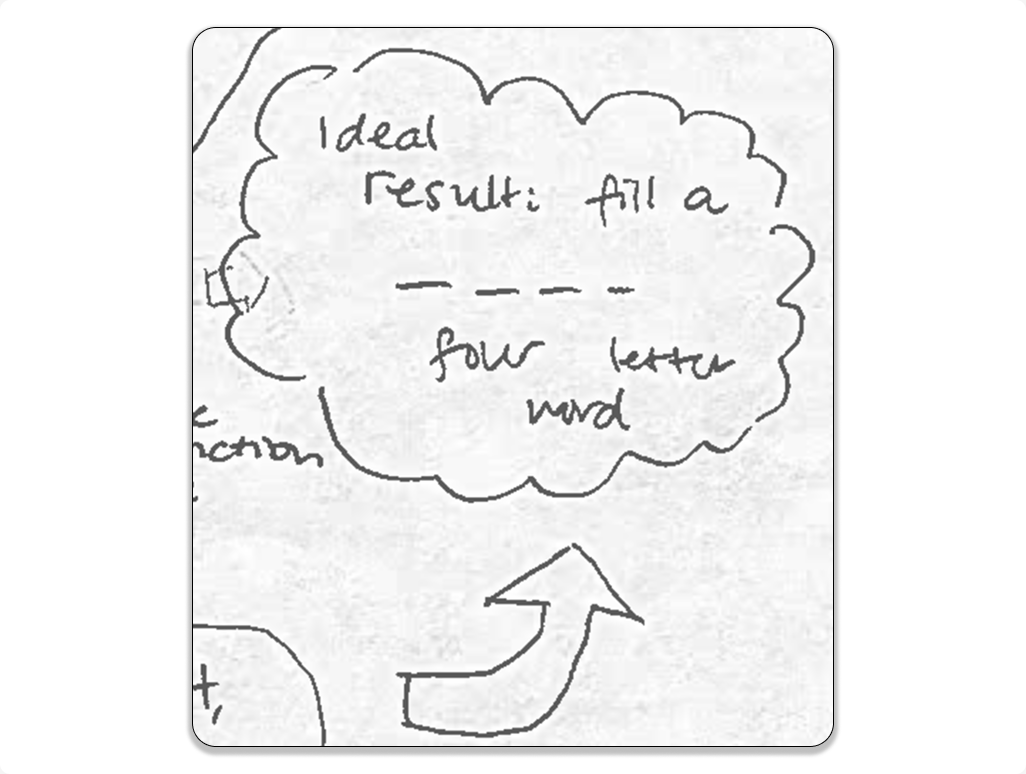}}
    \caption{Speech bubble interpreted by GenAI as a func.}
    \label{fig:ambig-b}
  \end{subfigure}
  \hfill
  \begin{subfigure}[b]{0.59\columnwidth}
    \centering
    \adjustbox{valign=m}{\includegraphics[width=\textwidth]{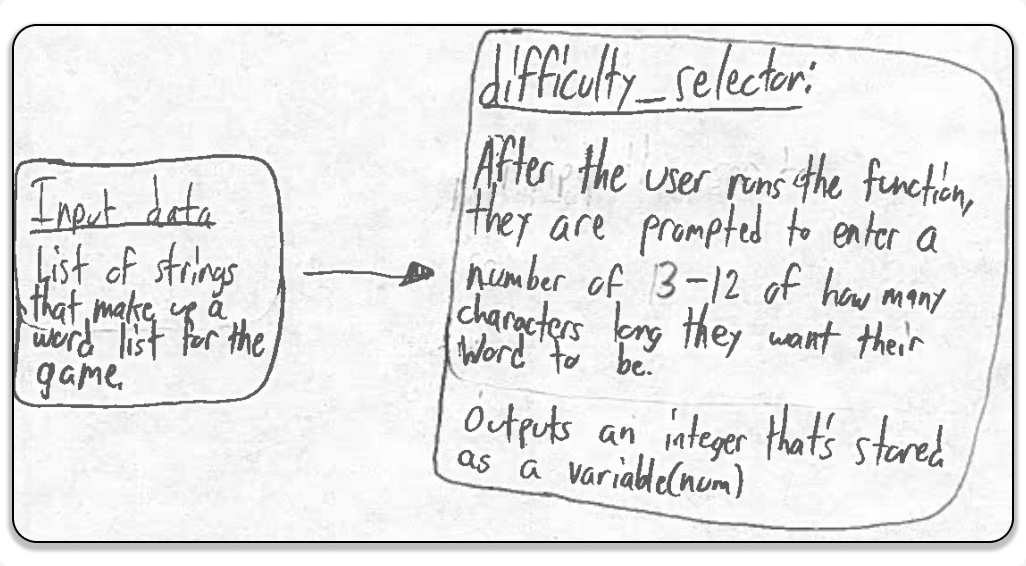}}
    \caption{An example of an input annotation that GenAI interpreted as a function.}
    \label{fig:ambig-c}
  \end{subfigure}
  \vspace{-0.25cm}
  \caption{Examples of ambiguous function definitions from student-generated
  decomposition diagrams.}\label{fig:annotation-errors}
  \vspace{-0.5cm}
\end{figure}

\noindent\textbf{Unclear Representations and Annotations} Though students
generally followed a predictable pattern of denoting functions through boxes or
clearly delineated groups of text, in many cases they added additional
annotations to help define the behavior of their decompositions. These
annotations often manifested as either external notes intended to clarify
function or program behavior (Figure~\ref{fig:ambig-c}). Misinterpretation of
function annotations---leading to generation of additional functions---was the
second most common reason the model produced more functions than identified by
humans.\\

\noindent\textbf{Ambiguity Between Function Names and Input} One primary
difference between functions identified by humans and those generated by GenAI
was the distinction between representations of functions and annotations
representing input. In many cases students represented program input of guesses,
words, or other inputs as a box with a label (Figure~\ref{fig:ambig-c}). This
could be considered a minor difference as, if such input representations had
been considered by human raters as independent functions, it likely would have
significantly improved between the human perceived number of functions in the
diagram and the number generated.\\

\noindent\textbf{Generated Helper Functions} In several cases where a
generated program contained more functions than anticipated by human raters,
this was due to ``helper functions'' containing functionality not explicitly
specified in the students' diagram. Examples include functions for displaying
game state, loading word banks, and introducing error handling.\\

\noindent\textbf{Failed Function Extractions} As might be expected,
a common source of error for programs containing fewer functions than
anticipated was failure to extract those functions. Upon closer inspection, this
was likely due to common causes of difficulty reading such diagrams: poor
handwriting, incomplete portions of diagrams, and other legibility issues. \\

\begin{tcolorbox}[
  colback=gray!10, colframe=gray!10,
  boxrule=0pt, arc=2pt,
  left=4pt, right=4pt, top=2pt, bottom=2pt,
  fontupper=\small, before skip=6pt, after skip=6pt
]
\textbf{Key Takeaway:} GenAI was largely effective at extracting functions from
decomposition diagrams, with mismatches appearing primarily attributable to
ambiguity in students' bespoke diagramming approaches or general legibility.
\end{tcolorbox}

\section{Discussion}
Our preliminary results are promising with respect to the efficacy of generating
code from diagrams. As such, this discussion focuses on the implications for
teaching and assessment in the context of computing education with respect to
some of the applications discussed in our motivating examples for visual
prompting listed in Section~\ref{sec:motivation}.

\subsection{Implications for Diagramming Guidelines}

Our results indicate that state-of-the-art GenAI models perform reasonably well
at extracting functions and composing programs from student-generated problem
decomposition diagrams, respectively. However, ambiguities can arise from how
functions and inputs are annotated and represented. Such ambiguities extend an
early concern of Dijkstra's~\cite{dijkstra2005foolishness} relating to natural
language programming---that the lack of formalism inherent to natural language
would make it an unreliable medium for constructing programs.
\begin{quote}
  \textit{The virtue of formal texts is that their manipulations, in order to be
    legitimate, need to satisfy only a few simple rules; they are, when you come
    to think of it, an amazingly effective tool for ruling out all sorts of
    nonsense that, when we use our native tongues, are almost impossible to
  avoid.}
\end{quote}
It appears, from our results, that this concern may apply to decomposition
diagrams. To reduce such ambiguities, it may be advisable to provide students
with clear guidelines on how to represent functions and their connections
within diagrams. This may not only be pedagogically
beneficial~\cite{reiser2018scaffolding, pea2018social}---by scaffolding the
diagramming task itself---but also improve interpretability for both GenAI
models and human readers. 

A semi-structured diagramming approach could offer the dual benefit of guiding
students in constructing effective diagrams while still allowing them to break
the rules at their discretion to explore how changes in representation affect
outcomes. This may strike a balance between a structured diagramming approach
and the full expressiveness made possible through sketching on paper.
Semi-structured approaches may preserve the proposed cognitive benefits of
visual representation (Section~\ref{sec:cognitive-load}) while reducing
interpretation errors on the part of the GenAI model.

\subsection{Implications for Developing Autograders}

Section 3.3.1 proposed that diagrams could function as visual prompts for
automated assessment where students submit a diagram, the model
generates code, and that code is evaluated against behavioral test cases. Our
findings suggest this vision is feasible, but they also surface several challenges
that should be addressed before such autograders can be deployed reliably.\\

\noindent \textbf{Challenge 1: How should generated code be evaluated?} In the
context of program decomposition diagrams, students define their own functions,
making traditional unit testing against specific signatures not viable for
assessing functional correctness. Instead, assessments must rely on end-to-end
behavioral testing by simulating input and verifying output. However, behavioral
correctness alone does not capture whether a student's diagram reflects sound
design principles. A program that passes all tests may still exhibit poor
structure, sidestepping the core learning objective. Static analysis techniques
could provide complementary feedback on decomposition quality, aligning with
frameworks that emphasize cohesion, reusability, and minimal information flow
between components~\cite{haldeman2025teaching}. Effective autograding will
likely require both approaches with behavioral testing to verify functionality and
static analysis to assess the quality of the design itself.\\

\noindent \textbf{Challenge 2: What counts as success when GenAI interpretation
of a diagram deviates from the actual diagram?} Our results showed that the
model often produced code that deviated from student intent—adding helper
functions, interpreting annotations as operations, or conflating inputs with
functions. In many cases, these deviations may still yield code that passes
behavioral tests. If an ambiguous diagram is interpreted in a way that happens
to work, has the student demonstrated decomposition skill? Future work should
address how these deviations impact the quality of the evaluation performed by
autograders that leverage a code generation based approach.\\

\noindent \textbf{Challenge 3: Is generation consistent enough for fair
grading?} Our study did not test whether the same diagram, submitted multiple
times, produces functionally equivalent code. Given the non-deterministic nature
of large language models, a student's grade could depend on which interpretation
the model happened to produce on a given run. Borrowing from work on EiPL
questions~\cite{smith2024prompting, smith2025redefining}, addressing this may
require multiple generation runs in order to ensure students descriptions are
unambiguous and therefore able to generate functionally equivalent code multiple
times.\\

\section{Limitations}
As with all work, this paper has limitations---particularly given its more
exploratory focus for the purpose of guiding future system design and
evaluations. Our evaluation relies on a single model---OpenAI's GPT 4.1---and a
single prompt, which, while state of the art, limits the generalizability of
our results to other models and approaches. Similarly, our evaluation was
limited to decomposition diagrams of a single task. Results may have differed
had an alternative task or a collection of tasks been used instead. As such,
future work should explore a broader range of tasks, models, and prompt
strategies to gain a fuller understanding of the design space for supporting
code generation from decomposition diagrams.

\section{Conclusion}
In this proposal and position paper, we have argued that problem decomposition
diagrams represent a promising and underexplored modality for Human-GenAI
collaboration as well as one that aligns with how students naturally sketch out
their ideas when learning to program. This approach opens the door to numerous
exciting possibilities, namely the use of sketches as a means of prompting for
code generation and, therefore, something of a novice friendly programming
interface in its own right. Future work in both extending and informing the use
of this approach should consider other approaches for providing students with
automated feedback on their diagrams and supporting students in prototyping
early designs of larger programs in formative contexts.

\section*{Acknowledgments}
Generative AI tools were used to assist with copy-editing and rewording during
the preparation of this manuscript as well as the creation of the teaser image. The authors reviewed all suggestions and
take full responsibility for the content.

\bibliographystyle{ACM-Reference-Format}
\bibliography{references}

\end{document}